\pdfoutput=1
\documentclass[journal=nalefd,manuscript=article,layout=twocolumn]{achemso}%
\usepackage{graphicx}%
\usepackage[separate-uncertainty = true]{siunitx}
\usepackage{physics}
\usepackage{amsmath}
\usepackage[T1]{fontenc}
\usepackage{hyperref}
\usepackage{pdfpages}
\usepackage[left]{lineno}

\title{Time-resolved single-particle x-ray scattering reveals electron-density gradients as coherent plasmonic-nanoparticle-oscillation source}
\author{Dominik Hoeing}
\affiliation{The Hamburg Centre for Ultrafast Imaging, Universität Hamburg, 22761 Hamburg, Germany}
\alsoaffiliation{Department of Chemistry, Universität Hamburg, 20146 Hamburg, Germany}
\author{Robert Salzwedel}
\affiliation{Institut für Theoretische Physik, Technische Universität Berlin, 10623 Berlin, Germany}
\author{Lena Worbs}
\affiliation{Center for Free-Electron Laser Science CFEL, Deutsches Elektronen-Synchrotron DESY, 22607 Hamburg, Germany}
\alsoaffiliation{Department of Physics, Universität Hamburg, 22761 Hamburg, Germany}
\author{Yulong Zhuang}
\affiliation{Max Planck Institut for the Structure and Dynamics of Matter, 22761 Hamburg Germany}
\author{Amit~K.~Samanta}
\affiliation{Center for Free-Electron Laser Science CFEL, Deutsches Elektronen-Synchrotron DESY, 22607 Hamburg, Germany}
\author{Jannik Lübke}
\affiliation{Center for Free-Electron Laser Science CFEL, Deutsches Elektronen-Synchrotron DESY, 22607 Hamburg, Germany}
\alsoaffiliation{The Hamburg Centre for Ultrafast Imaging, Universität Hamburg, 22761 Hamburg, Germany}
\alsoaffiliation{Department of Physics, Universität Hamburg, 22761 Hamburg, Germany}
\author{Armando D. Estillore}
\affiliation{Center for Free-Electron Laser Science CFEL, Deutsches Elektronen-Synchrotron DESY, 22607 Hamburg, Germany}
\author{Karol Dlugolecki}
\affiliation{Center for Free-Electron Laser Science CFEL, Deutsches Elektronen-Synchrotron DESY, 22607 Hamburg, Germany}
\author{Christopher Passow}
\affiliation{Deutsches Elektronen-Synchrotron DESY, 22607 Hamburg, Germany}
\author{Benjamin Erk}
\affiliation{Deutsches Elektronen-Synchrotron DESY, 22607 Hamburg, Germany}
\author{Nagitha Ekanayake}
\affiliation{Deutsches Elektronen-Synchrotron DESY, 22607 Hamburg, Germany}
\author{Daniel Ramm}
\affiliation{Deutsches Elektronen-Synchrotron DESY, 22607 Hamburg, Germany}
\author{Jonathan Correa}
\affiliation{Deutsches Elektronen-Synchrotron DESY, 22607 Hamburg, Germany}
\author{Christina C. Papadopoulou}
\affiliation{Deutsches Elektronen-Synchrotron DESY, 22607 Hamburg, Germany}
\author{Atia Tul Noor}
\affiliation{Deutsches Elektronen-Synchrotron DESY, 22607 Hamburg, Germany}
\author{Florian Schulz}
\affiliation{Department of Physics, Universität Hamburg, 22761 Hamburg, Germany}
\author{Malte Selig}
\affiliation{Institut für Theoretische Physik, Technische Universität Berlin, 10623 Berlin, Germany}
\author{Andreas Knorr}
\email{andreas.knorr@tu-berlin.de}
\affiliation{Institut für Theoretische Physik, Technische Universität Berlin, 10623 Berlin, Germany}
\author {Kartik Ayyer}\email{kartik.ayyer@mpsd.mpg.de}
\affiliation{The Hamburg Centre for Ultrafast Imaging, Universität Hamburg, 22761 Hamburg, Germany}
\alsoaffiliation{Max Planck Institut for the Structure and Dynamics of Matter, 22761 Hamburg Germany}
\author{Jochen Küpper}\email{jochen.kuepper@cfel.de}
\affiliation{Center for Free-Electron Laser Science CFEL, Deutsches Elektronen-Synchrotron DESY, 22607 Hamburg, Germany}
\alsoaffiliation{The Hamburg Centre for Ultrafast Imaging, Universität Hamburg, 22761 Hamburg, Germany}
\alsoaffiliation{Department of Physics, Universität Hamburg, 22761 Hamburg, Germany}
\alsoaffiliation{Department of Chemistry, Universität Hamburg, 20146 Hamburg, Germany}
\author {Holger Lange}\email{holger.lange@uni-hamburg.de}
\affiliation{The Hamburg Centre for Ultrafast Imaging, Universität Hamburg, 22761 Hamburg, Germany}
\alsoaffiliation{Department of Chemistry, Universität Hamburg, 20146 Hamburg, Germany}

\begin{document}
\maketitle
\twocolumn[
\begin{@twocolumnfalse}
\abstract{Dynamics of optically-excited plasmonic nanoparticles are presently understood as a series of sequential scattering events, involving thermalization processes after pulsed optical excitation. One important step is the initiation of nanoparticle breathing oscillations.
According to established experiments and models, these are caused by the statistical heat transfer from thermalized electrons to the lattice. An additional contribution by hot electron pressure has to be included to account for phase mismatches that arise from the lack of experimental data on the breathing onset.
We used optical transient-absorption spectroscopy and time-resolved single-particle x-ray-diffractive imaging to access the excited electron system and lattice. The time-resolved single-particle imaging data provided structural information directly on the onset of the breathing oscillation and confirmed the need for an additional excitation mechanism to thermal expansion, while the observed phase-dependence of the combined structural and optical data contrasted previous studies. Therefore, we developed a new model that reproduces all our experimental observations without using fit parameters. We identified optically-induced electron density gradients as the main driving source.\\
\\}
\end{@twocolumnfalse}
]
\pagebreak%

\section*{Introduction}
Plasmonics treats the unique optical excitation of metallic nanoparticles. The plasmon is a collective electron oscillation, associated with highly localized fields. It offers the ultimate spatial and temporal control over light, concentrating electromagnetic energy into nanoscale volumes. Applications range from catalysis and photovoltaics to sensing and quantum optics.\cite{reddy_determining_2020, tan_plasmonic_2017, zhou_hot_2021, celiksoy_intensity-based_2021, mueller_deep_2020} Presently, the details of the plasmon decay are intensely discussed. The current consensus is that very energetic "hot" electrons are generated, which are of interest for many applications. These non-equilibrium carriers thermalize via electron-electron scattering and then quickly couple to lattice phonons. The excited lattice finally dissipates the excess energy into the environment.\cite{boriskina_losses_2017,besteiro_fast_2019,brongersma_plasmon-induced_2015,hartland_optical_2011,linic_flow_2021} A detailed knowledge of these processes is the basis for controlling them for the respective applications. In particular the relaxation dynamics of hot electrons are of interest and as part of that, the coupling to coherent phonon oscillations producing a periodic change in particle dimension, radial "breathing" oscillations.\\

\begin{figure*}[!htpb]
   \includegraphics[width=\textwidth]{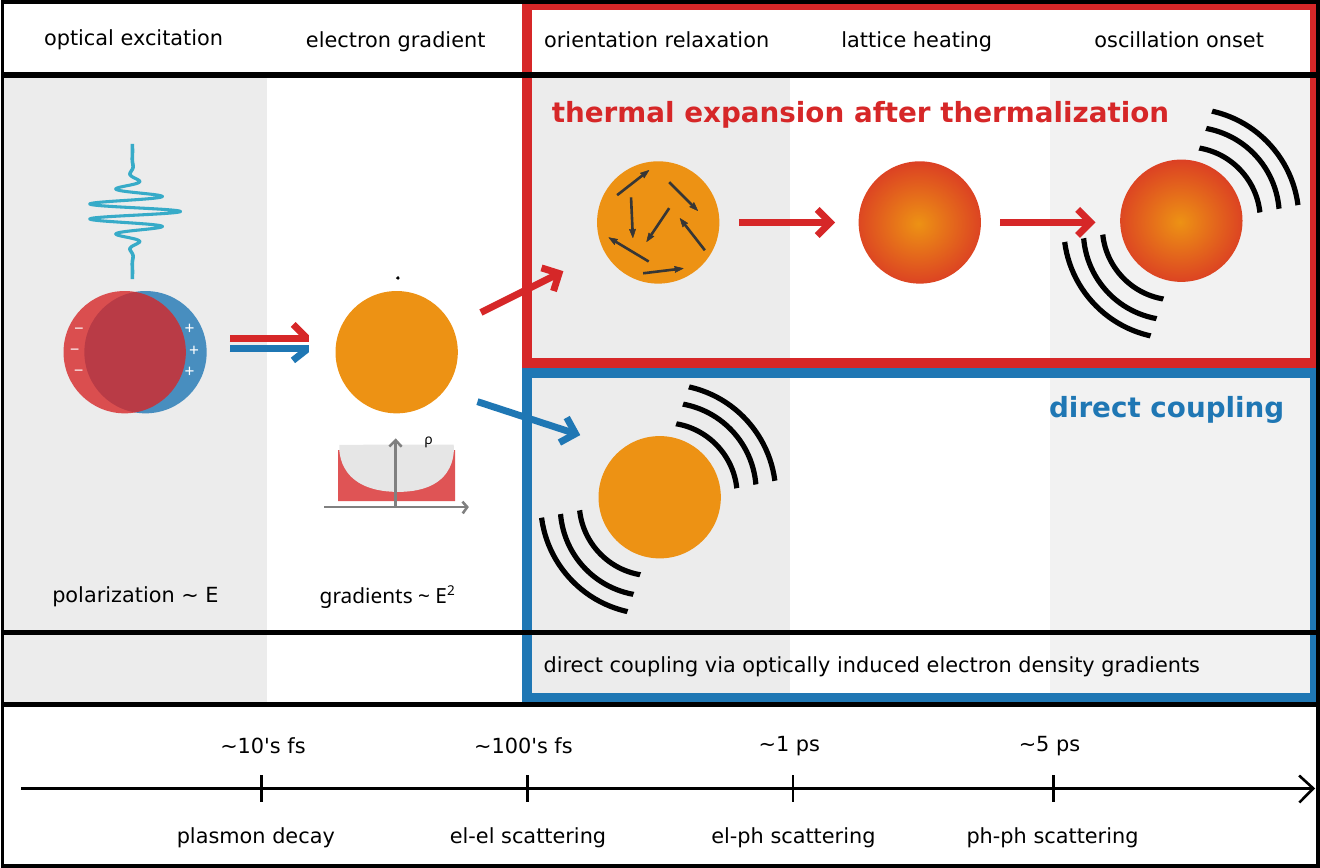}
   \caption{Schematic illustration of the relaxation dynamics in plasmonic nanoparticles: The
    optical pump causes a polarization of the electrons in the nanoparticle, which is accompanied by a density gradient. The red path illustrates the thermal-driving mechanism of the breathing oscillations: After Coulomb-driven orientation relaxation, electron-phonon scattering converts energy from the electrons to the lattice, which leads to an expansion of the particle. This initiates breathing oscillations. The blue path illustrates the coupling mechanism described in this work, the electron density gradients which directly couple to the coherent phonon mode, impulsively initiating breathing oscillations.}
    \label{fig:schema}
\end{figure*}

The temperature of the thermalized hot electron gas can be observed as a contrast in optical
transient-absorption (TA) experiments and conclusions on the lattice temperature can be made based on two-temperature models~\cite{hartland_optical_2011, brown_experimental_2017}. The breathing oscillations of plasmonic nanoparticles are then apparent as additional periodic contrast modulations in the TA spectra~\cite{arbouet_optical_2006,ahmed_understanding_2017,yu_damping_2013} and the observed dynamics is typically explained by mechanistic descriptions.\cite{hartland_optical_2011}
However, TA does not allow disentangling the precise temporal onset of the breathing oscillations from the contrast-dominating initial electron dynamics. Consequently, the details of the excitation were previously deduced indirectly. For example, matching the phase of the experimentally observed breathing oscillations and mechanistic descriptions required an additional impulsive excitation source to heat transfer, which was assigned to hot-electron pressure.~\cite{perner_observation_2000,wang_coherent_2020} In such descriptions, the different
time dependence of the hot electron pressure compared to the one for the lattice expansion affected the phase of the breathing oscillation. At high pump laser powers, where the hot electron pressure is assumed to be strongest, the change in phase was on the order of up to
45$^\circ$~\cite{perner_observation_2000}.
Today, x-ray free-electron lasers (XFELs) provide intense femtosecond x-ray pulses and allow for the time-resolved imaging of nanoparticle shapes, unaffected from electron temperature effects, down to femtosecond timescales~\cite{clark_ultrafast_2013, clark_imaging_2015, von_reppert_watching_2016, shin_ultrafast_2023}.
An emerging technique to conduct time-resolved XFEL experiments on nanoparticles is single-particle imaging (SPI)~\cite{Gorkhover:NatPhoton10:93, ayyer_3d_2021}, which combines a large number of diffraction patterns from individual particles into a diffraction volume, which is then inverted to a nanoparticle structure.\cite{ayyer_3d_2021} A major advantage of SPI is the combination of measuring single-particle images, allowing for correcting sample inhomogeneity, with the statistical robustness of a serial measurement. Using ultrashort x-ray pulses also allows to outrun potential radiation damage induced by the x-ray pulses themselves in a diffract-before-destruction process~\cite{neutze_potential_2000}.
Acquiring SPI data in a pump-probe fashion then provides a ``movie'' of the
nanoparticle dynamics, i.e., a direct measure of the nanoparticle's structural changes as a function of time.\\

Here, we exploited XFEL transient small-angle-x-ray-scattering in a single-particle imaging scheme (tSAX-SPI) and TA spectroscopy to unravel the size and electron temperature of gold nanoparticles (AuNP) as a function of time after optical excitation. The combined SPI and TA experimental observations could not be fully explained within established models. The particle expansion started already with the optical excitation pulse and thus occured in the kinetic limit of the electron gas excitation, see \autoref{fig:schema}.
This directly confirms the need for an impulsive-excitation source, which is not the lattice temperature rise. However, additional power-dependent TA experiments showed no phase change, contrary to hot-electron pressure based expectations.
Our experimental findings were rationalized by calculations based on a combination of the previous, well established model and an additional new source for coherent AuNP size oscillations yielding a consistent picture of all observed aspects of the breathing excitation.\cite{salzwedel_theory_2023} Two source terms for the lattice dynamics emerge from the calculations: The optically-induced spatial gradients of the electron density act as the main initial driving source for the breathing oscillations, accompanied by time-delayed thermal contributions. The dominance of the direct coupling between the displaced electrons and the lattice can have important consequences for applications harnessing hot electrons~\cite{sytwu_driving_2021,zhang_surface-plasmon-driven_2018,tan_plasmonic_2017,zhou_hot_2021,baffou_applications_2020} and for the general understanding of nanoscale metal dynamics.
\pagebreak
\section*{Results}
\paragraph*{Pump-probe spectroscopy of gold nanoparticles\\}

\begin{figure*}[!htpb]
    \includegraphics[width=\textwidth]{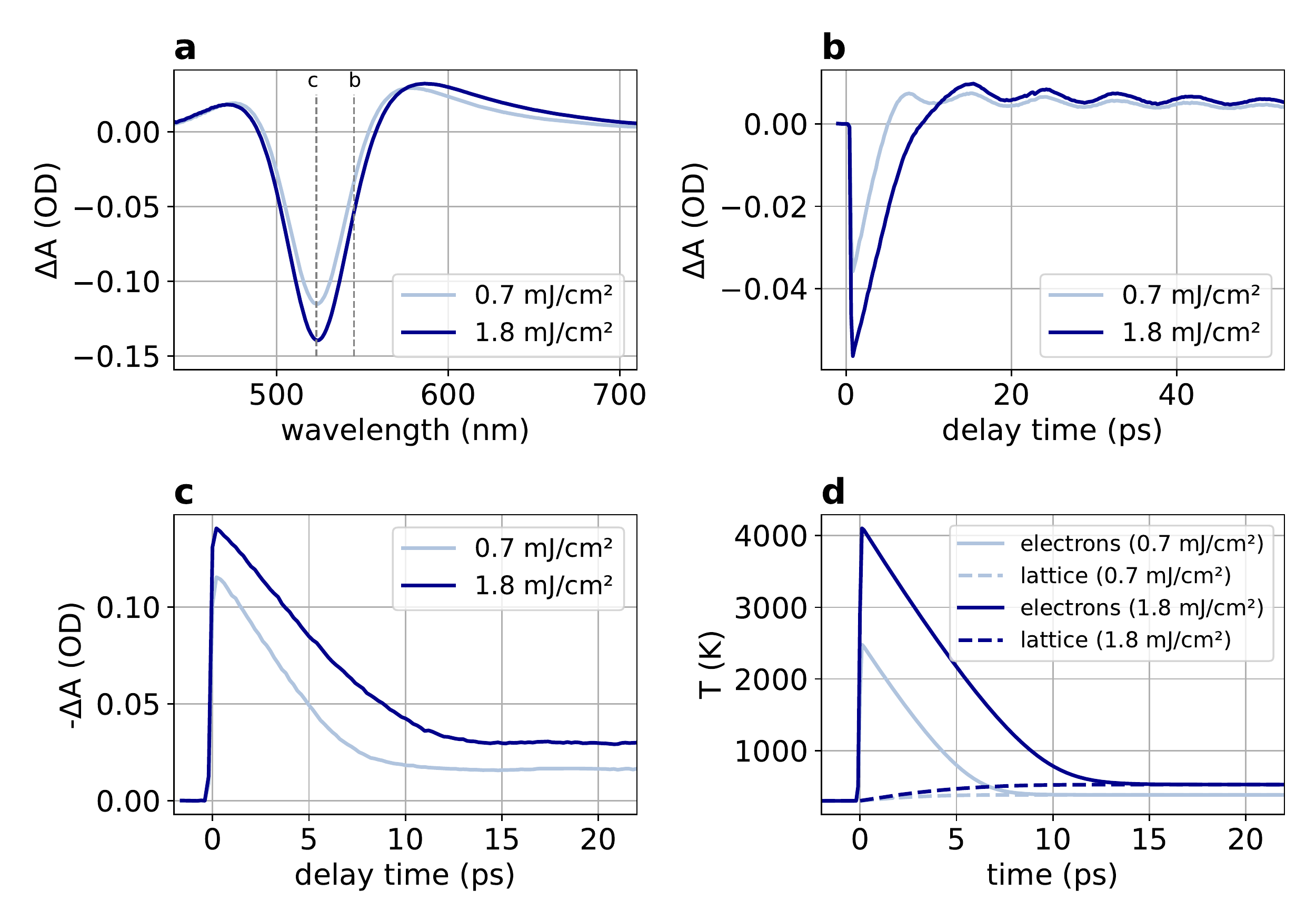}
    \caption{(a) TA spectra at the delay time of maximum contrast, i.e., \SI{500}{\fs}. The vertical dashed lines mark the probed wavelengths for b) and c). (b) TA kinetics probed at the bleach's long-wavelength shoulder (\SI{545}{\nm}). (c) TA kinetics probed at the bleach centre-of-mass (ca.~\SI{525}{\nm}). (d) Electron and lattice temperatures according to the two-temperature model (see Supporting Information Section 4).}
    \label{fig:TAresults}
\end{figure*}

We recorded the TA spectra of \SI{27}{\nm}, mono-crystalline, spherical AuNPs, stabilized in water with cetyltrimethylammonium chloride (see Supporting Information Section 1 for sample
characterization). The excitation of the AuNPs with femtosecond laser pulses at a central wavelength of \SI{400}{\nm} led to TA spectra shown in Figure~\ref{fig:TAresults}a. The shape of the spectra - a central bleach at the plasmon wavelength with positive sidebands - resulted from the change of the complex dielectric function of gold at elevated electron temperatures\cite{brown_ab_2016}. AuNP size changes due to the breathing oscillations led to periodic modulations of the plasmon center wavelength, observable on the long-wavelength side of the TA spectrum (Figure~\ref{fig:TAresults}b, Ref. \citenum{hartland_optical_2011}). The temporal onset of the breathing oscillations were not visible, because the signal produced by hot electrons dominates the overall contrast. Modeling the evolution of the electron temperature with a two-temperature model reproduced the TA bleach kinetics (Figure~\ref{fig:TAresults}c-d). We conclude that the electron cooling time is excitation-fluence dependent and occurs on timescales of tens of picoseconds for the employed excitation conditions. As heat losses to the environment occur on much longer timescales, the lattice temperature increase can be assumed to coincide with the electron temperature decrease,
Figure~\ref{fig:TAresults}d.

\paragraph*{Transient small-angle-x-ray-scattering\\}
To assess the AuNP sizes in a direct fashion, we performed tSAX-SPI at the free-electron laser
FLASH. Individual AuNPs were sequentially injected into the x-ray beam and excited with fs laser
pulses of the same wavelength (\SI{400}{nm}) and fluence (\SIlist[list-units=single]{0.7; 1.8}{mJ/cm^2}) as for the TA experiments. The AuNP diameter changes were then probed by time-delayed ultrashort x-ray pulses. The serial nature of the measurement enabled the analysis of only those patterns which came from highly spherical particles. Any patterns which showed significant ellipticity were rejected (see Supporting Information for experimental and analysis details).

\begin{figure*}
    \centering
    \includegraphics[width = \textwidth]{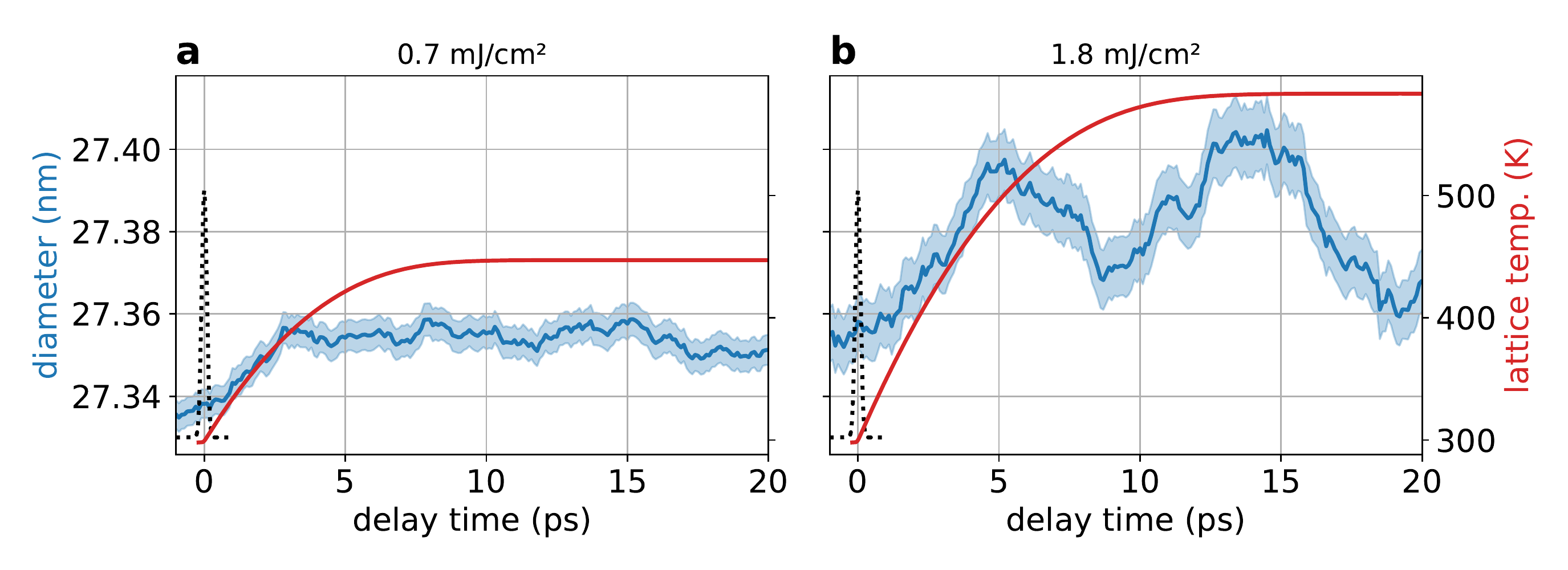}
    \caption{(blue) AuNP diameter according to tSAX-SPI experiment and (red) corresponding lattice temperatures according to a two-temperature model for two excitation fluences: (a) \SI{0.7}{mJ/cm^2} and (b) \SI{1.8}{mJ/cm^2}. The shaded area is the AuNP diameter standard deviation. The envelope of the optical excitation pulse is indicated as dotted line.}
    \label{fig:flashresults}
\end{figure*}

For both employed pump fluences we observed an immediate AuNP diameter expansion with photoexcitation  (Figure~\ref{fig:flashresults}). An oscillation half cycle already occurred prior to the first oscillation being visible in the TA dynamics (cf. Figure~\ref{fig:TAresults}b). The magnitude of the diameter increase was excitation-power dependent and the breathing oscillation was more clearly visible for the higher pump fluence of \SI{1.8}{mJ/cm^2}. In Figure~\ref{fig:flashresults}, we compare the experimental diameter oscillation with the lattice temperature obtained from the two-temperature model. The particle diameter reaches its maximum at a delay time of around \SI{3}{\ps} (\SI{5}{\ps}) for a pump fluence of \SI{0.7}{mJ/cm^2} (\SI{1.8}{mJ/cm^2}) while the lattice temperature reaches its maximum at around \SI{14}{\ps} (\SI{17}{\ps}). Considering the timescales, the thermal lattice expansion cannot be the main oscillation-driving impulsive source, since its rise time is longer than the breathing oscillation period.
\paragraph*{Theory of electronic and thermal sources to the oscillation\\}
To understand the origin of the breathing oscillations, we set up a fundamental microscopic model which extends earlier calculations in Ref.~\citenum{salzwedel_theory_2023} to two-band models. We started with the definition of the lattice displacement at position $\vb{r}$ and time $t$ as observable for the AuNP breathing oscillations, cf. S1:
\begin{equation}
\begin{aligned}
\vb{u}_\beta(\vb{r},t)=
	\sum_{\vb{q}}
	\sqrt{\frac{\hbar}{2mN\omega_{\beta,\vb{q}}}}
	\vb{A}_{\beta,\vb{q}}
	e^{i\vb{q}\cdot\vb{r}}\\
	\expval{
	b_{\beta,-\vb{q}}^\dagger(t)
	+b_{\beta,\vb{q}}(t)
	},
\end{aligned}
\end{equation}
\noindent with the reduced Planck constant $\hbar$, phonon dispersion $\omega_{\beta,\vb{q}}$ and phonon annihilation (creation) operators $b^{(\dagger)}_{\beta,\vb{q}}$ with momentum $\vb{q}$ and branch $\beta$. Further, $m$ accounts for an effective mass of the unit cell, $\mathbf{A}_{\beta , \mathbf{q}}$ for the polarization vector and $N$ for the number of unit cells in the crystal. Focusing on the longitudinal acoustic (LA) phonon, $\beta = LA$, we calculated the equation of motion for the lattice displacement. The underlying Hamiltonian includes the electron-phonon interaction for a two-band model of the AuNP electrons including interband transitions as well as phonon-phonon interaction. (see  Supporting Information Section 3). We obtain:
\begin{equation}
\begin{aligned}
    \qty[
        \partial_t^2 + 2\gamma_p\partial_t -c_{LA}^2\nabla^2
    ]
    \vb{u}(\vb{r},t)
    =\\
    \boldsymbol{\xi} \qty[ T(t)-T_{0}]
    + \zeta\,\nabla\rho^c(\vb{r},t). \label{eq:displacement}
\end{aligned}
\end{equation}
The left-hand side accounts for a damped oscillator equation for the lattice displacement with the longitudinal velocity of sound $c_{LA}$ and the damping rate $\gamma_p$. The right-hand side describes the sources of the lattice displacement: The thermal lattice expansion arises from the phonon-phonon interaction and accounts for the displacement of the equilibrium position as a function of the lattice temperature $T(t)$ with the coupling constant $\boldsymbol{\xi}$, cf. Figure~\ref{fig:schema} (red path). $T(t)$ was obtained by applying a two-temperature model (cf. Figure \ref{fig:TAresults}d and  Supporting Information Section 3) and $T_0$ is the initial equilibrium temperature. Our microscopic approach revealed a second source term arising from the electron-phonon interaction. On the right-hand side of Eq. (\ref{eq:displacement}), the second term describes the excitation of coherent oscillations resulting from the optically-induced temporal buildup of the spatial gradient of the electron charge density $\nabla \rho^c(\vb{r},t)$ in the conduction band of the AuNP, cf. Fig~\ref{fig:schema} (blue path) and  Supporting Information Section 3.\\

To connect the lattice diplacement with the optical excitation, we calculated the equation of motion for the conduction band electron density $\rho^c (\vb{r},t)$ \cite{salzwedel_theory_2023,grad_kinetic_1949,gardner_quantum_1994,cai_quantum_2012} incorporating the electron-light interaction and the electron-phonon interaction from the same Hamiltonian used above.
In the limiting case of a short interband dephasing time compared to the pulse width, a rate equation for the electron density can be derived and expanded in orders of the exciting electric field, i.e. $\rho = \rho^0 + \rho^1 + \rho^2 + \mathcal{O} (E^3)$. We find that the densities are determined by a source term that scales with the electric field intensity:
\begin{equation}
    \partial_t \rho_2^c(\vb{r},t)
    = \frac{2e\epsilon_0}{\hbar}\abs{\vb{E}(\vb{r},t)}^2\Im{\chi^{\text{inter}}(\omega_{\text{opt}})}, \label{eq:conti}
\end{equation}
where $\chi^{\text{inter}}(\omega)$ is the interband contribution to the electric susceptibility of the AuNP, $e$ is the elementary charge and $\epsilon_0$ is the vacuum dielectric constant. Note that spatial gradients of the electron density are induced via the boundary conditions of Maxwell's equations for the self-consistently calculated electric field $\vb{E}(\vb{r},t)$ showing amplification effects at the surface of the AuNP.\\

The coupled set of equations was solved by expansion into the vibrational eigenmodes of an oscillating sphere (Refs. \citenum{lamb_vibrations_1881}, Supporting Information Section 3).
Both, thermal and electron density source terms in Eq.~\ref{eq:displacement} are indirectly initiated by the exciting optical field $\mathbf{E} (\mathbf{r},t)$.
However, the thermal contribution is delayed via the temporal buildup from non-equilibrium, whereas the electron density gradient is directly induced by the optical field.
We found that the electron density gradient $\nabla \rho^{c}_2(\vb{r},t)$ within the AuNP is the main source of the lattice displacement, since thermal expansion occurs with a time delay of a few ps and thus cannot explain the observed onset of the lattice displacement.
\begin{figure*}
     \centering
     \includegraphics[width= \textwidth]{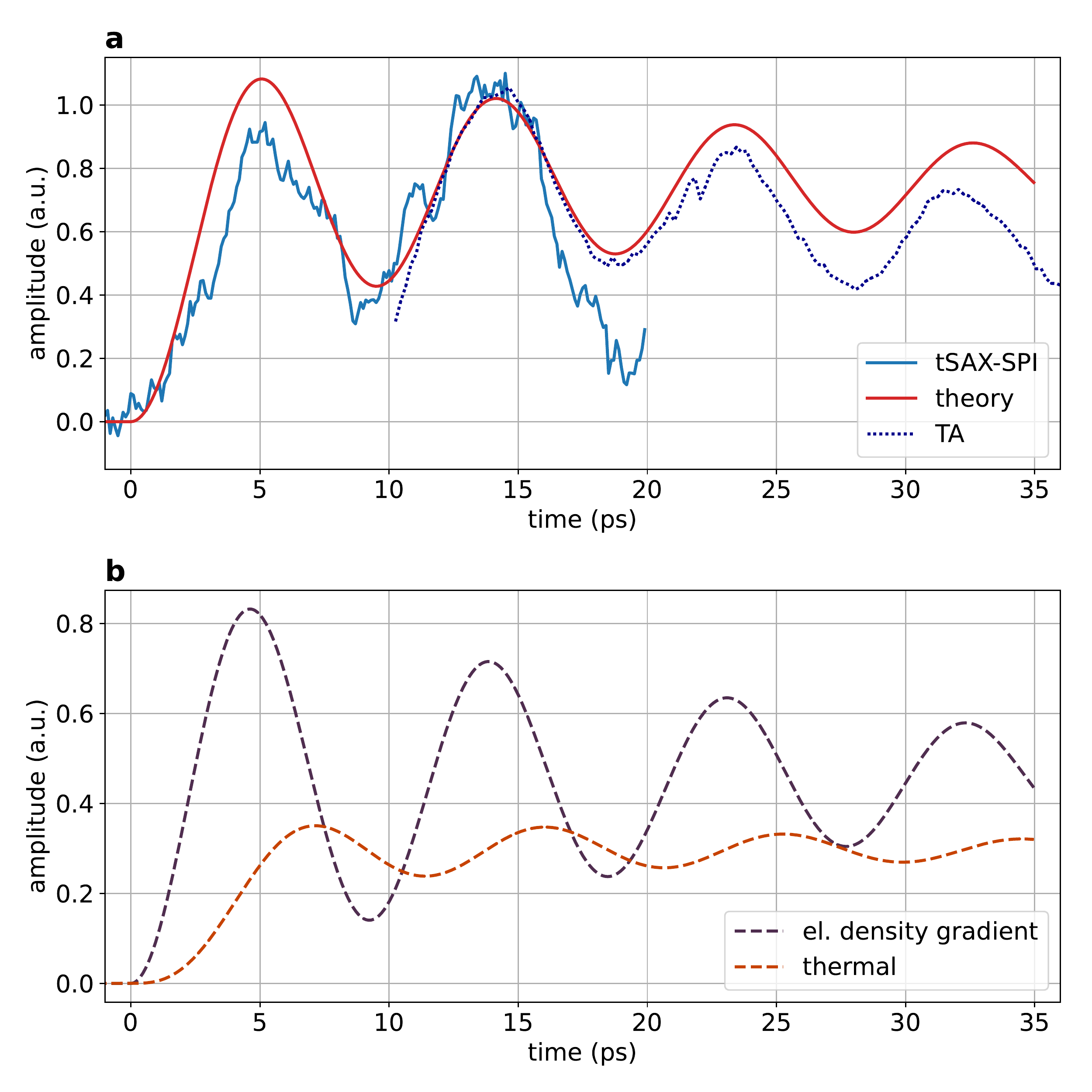}
 	    \caption{Experiment-theory comparison: (a), Particle size according to tSAX-SPI, TA kinetics at long-wavelength shoulder of the LSPR (the data is plotted for times after the initial electron cooling for better visibility and scaling.) and theoretical prediction of the particle diameter. The amplitudes were rescaled for easier comparability. (b), Relative contributions of electron density gradient and thermal expansion to the full theoretical prediction of the breathing oscillation.
 	    }
 	    \label{fig:theo1}
\end{figure*}
Figure~\ref{fig:theo1}a shows the calculated evolution of the lattice displacement in comparison to the experimental data. We found an excellent agreement between experiment and theory. In particular, the onset -- and thus the phase -- of the breathing oscillations is well reproduced by our formalism. Figure~\ref{fig:theo1}b illustrates the temporal evolution of the two contributions (thermal and electronic) to the displacement.
The contribution due to the spatial electron gradient starts immediately after the optical pulse since the electron density gradient scales with the electric field intensity, Eq.~\ref{eq:conti}. On the timescales of the breathing oscillation and the electron cooling, the development of the electron density gradient and the phonon excitation are simultaneous. The thermal expansion of the lattice also contributes to the excitation of the breathing oscillation, but to a lesser extent and on a slower timescale.\\

\begin{figure*}[htpb!]
    \centering
    \includegraphics[width = 0.7\linewidth]{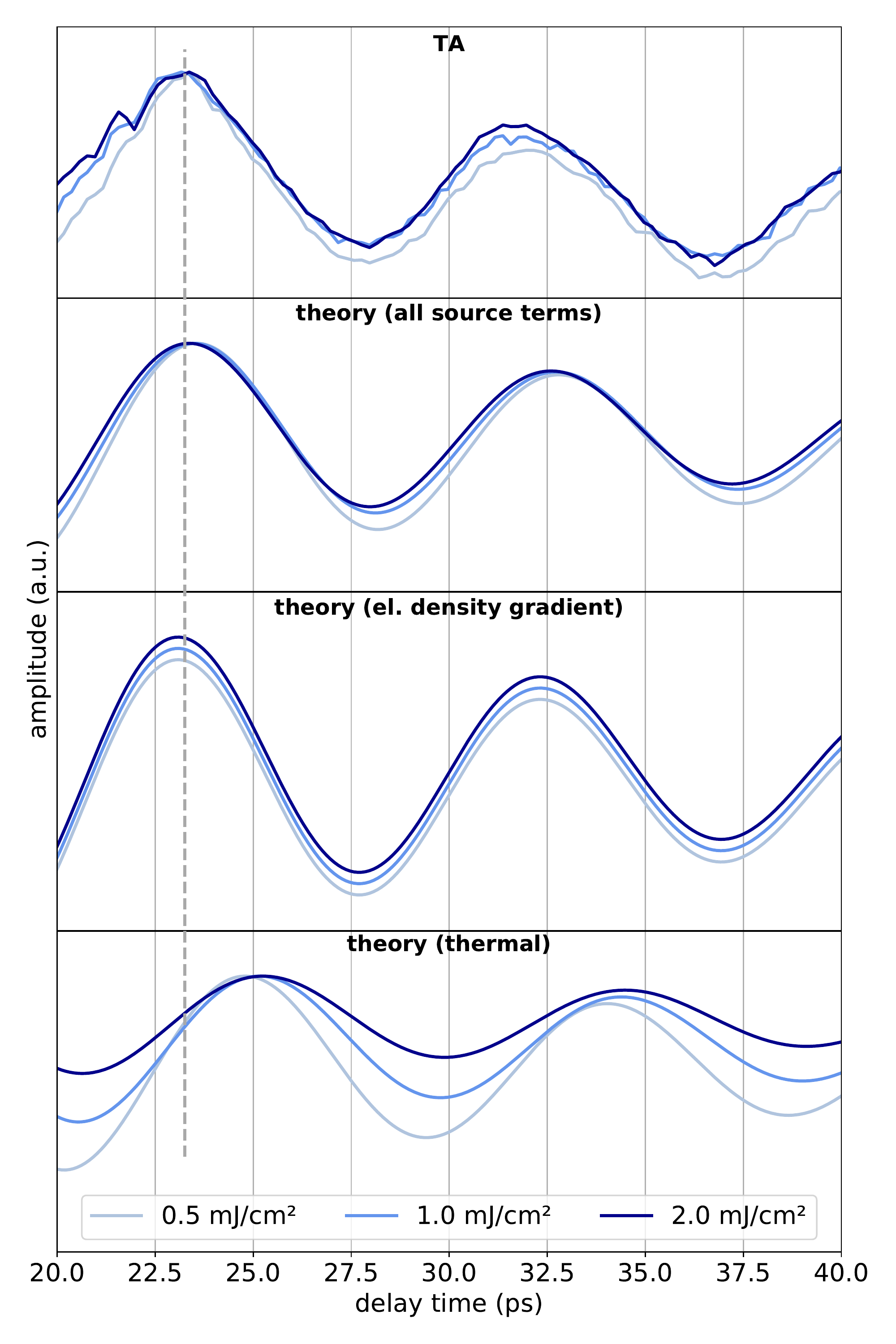}
    \caption{Experiment-theory comparison: Phase dependence on excitation fluence. TA data recorded for increasing pump fluences. "theory (all source terms)" includes both source terms in Eq.~(\ref{eq:displacement}) while "theory (el. density gradient)" and "theory (thermal)" are expectation values if only one of the sources is considered, respectively. Theoretical data for the oscillation due to the electron density gradients include an offset for better visibility. All data were normalized to their maximum value. The maximum of the TA data is indicated by a dashed line for better visibility of the phase difference.}
    \label{fig:TAphase}
\end{figure*}

This is apparent also in the fluence-dependence: As the duration of the lattice temperature increase depends on the initial electron gas temperature (deposited energy) an excitation-fluence-dependent phase enters the temperature-based models. \cite{voisin_time-resolved_2000,hartland_optical_2011}. Figure \ref{fig:TAphase}a displays TA data for the first resolvable breathing oscillations for increasing fluences.
The fluence-dependent experimental data lacks any phase dependence.
In the microscopic theory, the source term is dominated by electronic contributions, which show no fluence-dependent phase shifts, in agreement with the experiment.
As expected, the thermal component shows a fluence-dependent phase, supporting the dominant relative contribution of the electronic excitation to the breathing oscillation.

\section*{Conclusion}
Overall, we recorded time-resolved structural data of the AuNP breathing mode onset using single-particle imaging, which clearly confirm the necessity of two excitation sources for the breathing oscillations in AuNP, as was previously indicated indirectly from phase discrepancies. The combined time-resolved structural and optical data required the addition of direct interactions of the electronic system with coherent phonons to the excitation and phonon-lattice-coupling model, which were not only a necessary addition to the thermal driving terms but on short time scales indeed the dominant source term.
Our theory quantitatively explains all experimental findings and provides general access to plasmon-lattice interaction. The concurrency of initial electron and lattice dynamics, caused by the immediate coupling of the optically-induced electron-density gradient and the breathing oscillation, might also have strong implications on energy transformations involving plasmonic hot carriers.

\section*{Data Availability}
All data needed to evaluate the conclusions in the paper are present in the paper or the
supplementary materials. Primary data is available from the corresponding authors upon reasonable request.\\

\section*{Supporting Information}
Supporting Information with details on the following topics is available alongside
this manuscript:
\begin{itemize}
\item Methods
\item Data analysis of transient small-angle x-ray-scattering patterns
\item Theory of electric and thermal sources to the oscillation
\item Two-temperature model
\end{itemize}

\section*{Author Contributions} D.H.\ performed the TA experiments and synthesized the AuNPs together with F.S.; R.S.\ and M.S.\ developed the microscopic model, supervised by A.K.; L.W., J.L., A.D.E., K.D., and A.K.S. designed and performed the sample injection at the FEL. J.K.\ and A.K.S.\ devised and developed the XFEL experimental approach and designed and coordinated the experiment. The FEL experiment was set up and carried out by A.D.E., A.K.S., A.T.N., B.E., C.C.P., C.P., D.H.,
D.R., F.S., H.L., J.C., J.K., J.L., K.A., K.D., L.W., N.E., and Y.Z.; Y.Z.\ and K.A.\ analyzed the FEL data. K.A., J.K., and H.L.\ conceived the project and H.L.\ coordinated the project and drafted the manuscript together with D.H.. All authors contributed to the writing of the manuscript and the supporting material.

\section*{Acknowledgments}
This work was supported by the German Research Foundation (DFG) by the federal cluster of excellence ``Advanced Imaging of Matter'' (EXC~2056, ID~390715994). We acknowledge the Deutsches Elektronen-Synchtrotron DESY, a member of the Helmholtz Association (HGF), for financial support as well as for the provision of experimental facilities and the Maxwell computational resources. Beamtime was allocated for proposal F-20190741. We acknowledge the Max Planck Society for funding the development and the initial operation of the CAMP end-station within the Max Planck Advanced Study Group at CFEL and for providing this equipment for CAMP@FLASH. The installation of CAMP@FLASH was partially funded by the BMBF grants 05K10KT2, 05K13KT2, 05K16KT3 and 05K10KTB from FSP-302. J.K.\ acknowledges support by the European Research Council through the Consolidator Grant COMOTION (614507). H.L., R.S., and M.S.\ acknowledge funding by the DFG (432266622).
\bibliography{main}



\section*{Supplementary material}

# Supporting Information:

# Time-resolved single-particle x-ray scattering reveals electron-density gradients as coherent plasmonic-nanoparticle-oscillation source

Dominik Hoeing,[†,‡] Robert Salzwedel,[¶] Lena Worbs,[§,∥] Yulong Zhuang,[⊥] Amit K. Samanta,[§] Jannik Lübke,[§,†,∥] Armando D. Estillore,[§] Karol Dlugolecki,[§] Christopher Passow,[#] Benjamin Erk,[#] Nagitha Ekanayake,[#] Daniel Ramm,[#] Jonathan Correa,[#] Christina C. Papadopoulou,[#] Atia Tul Noor,[#] Florian Schulz,[∥] Malte Selig,[¶] Andreas Knorr,[*,¶] Kartik Ayyer,[*,†,⊥] Jochen Küpper,[*,§,†,∥,‡] and Holger Lange[*,†,‡]

[†]*The Hamburg Centre for Ultrafast Imaging, Universität Hamburg, 22761 Hamburg, Germany*
[‡]*Department of Chemistry, Universität Hamburg, 20146 Hamburg, Germany*
[¶]*Institut für Theoretische Physik, Technische Universität Berlin, 10623 Berlin, Germany*
[§]*Center for Free-Electron Laser Science CFEL, Deutsches Elektronen-Synchrotron DESY, 22607 Hamburg, Germany*
[∥]*Department of Physics, Universität Hamburg, 22761 Hamburg, Germany*
[⊥]*Max Planck Institut for the Structure and Dynamics of Matter, 22761 Hamburg Germany*
[#]*Deutsches Elektronen-Synchrotron DESY, 22607 Hamburg, Germany*

E-mail: andreas.knorr@tu-berlin.de; kartik.ayyer@mpsd.mpg.de; jochen.kuepper@cfel.de; holger.lange@uni-hamburg.de

1

# Methods

## Particle Synthesis

Tetrachloroauric(III) acid (HAuCl$_4$, $\geq 99.9\%$ trace metals basis), hexadecyltrimethylammonium bromide (CTAB, $\geq 98\%$) and chloride (CTAC, $\geq 98\%$), l-ascorbic acid (AA, reagent grade) and sodium borohydride (NaBH$_4$, $\geq 98\%$) were purchased from Sigma-Aldrich (USA). All reagents were used as received. Ultrapure water was used for all procedures.

Spherical gold nanoparticles (AuNPs) were synthesized according to a modified protocol by Zheng et al.[1,2] The protocol starts by producing Au clusters, followed by two growth steps. Au clusters were synthesized by adding 600 µL of an aqueous solution of NaBH$_4$ (10 mM) to a mixture of 0.1 ml HAuCl$_4$ (0.25 mM) and 10 mL CTAB (100 mM) via one-shot injection. The solution was rapidly stirred at 700 rpm during injection and for an additional 3 minutes. The solution was then kept undisturbed for 3 hours yielding Au clusters stabilized by CTAB.

The clusters were grown to 10.6 nm AuNPs by a one-shot-injection of the seed solution (50 µL) into a mixture of HAuCl$_4$ (2 ml, 0.5 mM), CTAC (2 ml, 200 mM) and ascorbic acid (AA, 1.5 ml, 100 mM). The solution was stirred for 15 minutes at 300 rpm and then washed twice by centrifugation (20 000 G, 1 h and 30 min, respectively) and replacement of the supernatant with a fresh CTAC solution (10 ml, 20 mM). The particle sizes were then increased again by a second growth step: A solution of the 10.6 nm AuNPs (3.6 mL) was mixed with a CTAC solution (120 ml, 100 mM). AA (780 µL, 100 mM) was added and after 60 s, HAuCl$_4$ (120 ml, 0.5 mM) was added dropwise within 1 hour with a syringe pump at a constant rate of 120 ml/h. The solution was stirred throughout the addition of AA and HAuCl$_4$ and for an additional 10 minutes at 300 rpm. The solution was washed and the concentration increased by centrifugation and replacement of the supernatant with sequentially less volume of fresh CTAC solution (5000 G, 2x 40 min and 3x 20 min). After the second centrifugation and redispersion step, the solution was transfered from 40 ml to (several) 1.5 ml centrifugation tubes. The final pellet was redispersed in 6 ml CTAC (20 mM) yielding a AuNP concentration of 12.9 nM ($7.76 \times 10^{12}$ particles/ml).

# Sample characterization

**UV-Vis spectroscopy**

Absorption spectra were recorded using a Varian Cary 50 spectrometer. The concentrated AuNP dispersion was diluted by a factor of 1:100 and its absorbance at 450 nm was used to determine the concentration according to Haiss et al.[3] The measured spectrum is plotted in Fig. S1. The shape of the plasmon-related absorption around 525 nm confirms isolated spherical AuNPs.

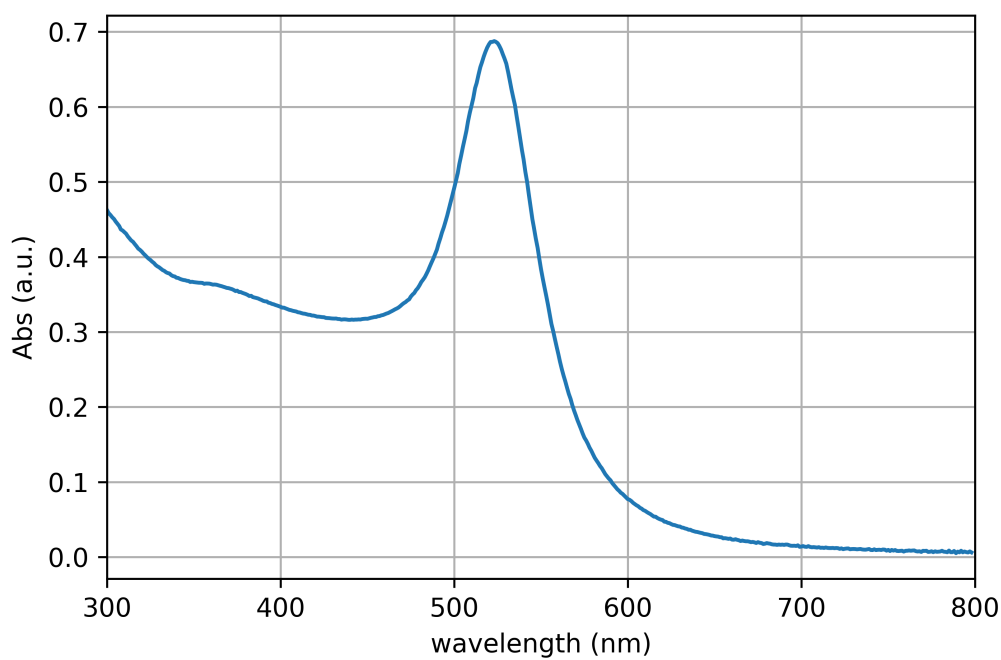

Fig. S1: UV-Vis spectrum of the employed AuNP solution.

**Transmission electron microscopy**

Transmission electron miscroscopy (TEM) was used to determine the AuNP size distributions. Because of the TEM contrast depending on the material's atomic number, particle sizes without contributions of the ligand layer are obtained, which is advantageous for a comparison with the results from the X-ray scattering experiment. According to the TEM analysis, the AuNPs used in the experiments were 27.1 nm in diameter with a standard deviation of 0.8 nm. Representative TEM images and the size histogram are displayed in Fig. S2.

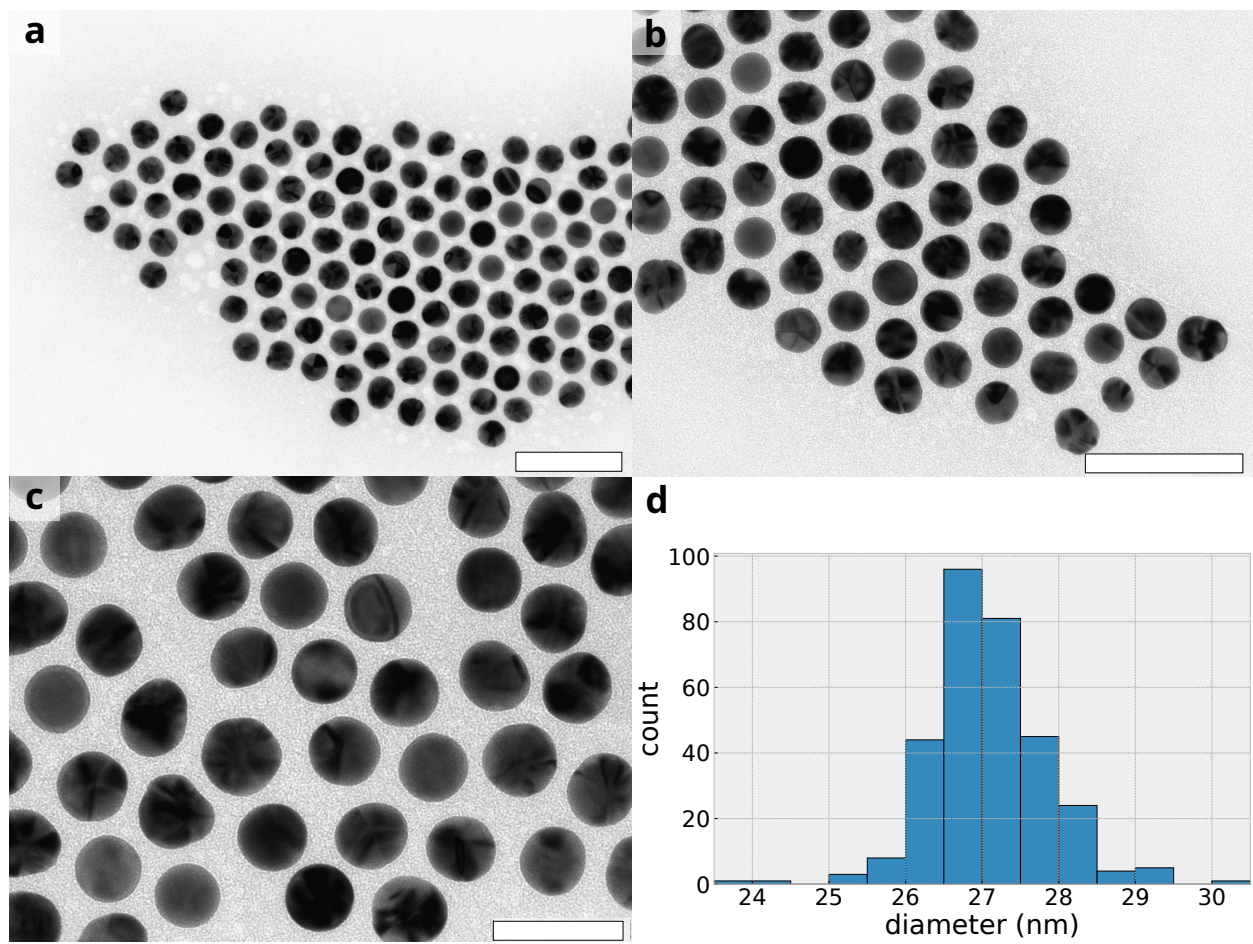

Fig. S2: **TEM analysis of gold nanoparticles. a-c,** Representative TEM images and **d**, size histogram of the employed AuNPs. The scale bars measure 200 nm (a), 100 nm (b) and 50 nm (c). The average particle diameter is (27.1 ± 0.8) nm.

**Transient Absorption Spectroscopy**

Ultrashort laser pulses with a center wavelength of 800 nm were generated in a chirped pulse amplifier (Spectra Physics Spitfire) at a repition rate of 1 kHz. The amplifier was seeded by a Ti:sapphire oscillator (Spectra Physics MaiTai) and pumped by a Nd:YBG laser (Spectra Physics Empower). The output of the regenerative amplifier is split into two arms, which are compressed individually. The pulse length was 35 fs as determined by autocorrelation.

One arm was sent through a BBO crystal to produce pump pulses with a center wavelength of 400 nm. Every second pump pulse was blocked by a chopper running at 500 Hz for referencing every pumped spectrum with a ground-state spectrum. The pump power was set to 0.7 and 1.8 mJ/cm$^2$ using a variable neutral density filter. Additionally, to study the phase-dependence of the breathing oscillation, 0.5, 1.0 and 2.0 mJ/cm$^2$ were applied. Because of the use of transmittive optical elements, an induced chirp is present, which results in an increase in pulse duration. However, due to the limited number of optical elements we expect the pump pulses to be shorter than 150 fs.

The second (probe) arm was aligned through a delay stage and onto a sapphire crystal for white-light generation in the transient absorption spectrometer (Ultrafast Systems Helios). The white-light transmitted through the sample was focused on a UV-Vis spectrometer with a 1024 pixel CMOS sensor.

The TA data was corrected for optical chirp and scattering around the pump wavelength. To obtain the e-ph time constants, the local minimum of the bleach signal at each delay time was determined using a gaussian function. Note that the TA minimum blueshifts within the first few picoseconds and therefore analyzing the bleach signal at a fixed probe wavelength can produce errors.[4] The obtained minimum of the bleach intensity vs. delay time was then normalized and fitted with a bi-exponential decay function. The time constants of the decay are interpreted as coupling times:

$$\Delta A(t) = a \cdot \exp\left(-\frac{t}{\tau_{e-ph}}\right) + b \cdot \exp\left(-\frac{t}{\tau_{ph-ph}}\right),$$

where $\tau_{e-ph}$ is the electron-phonon coupling time $\tau_{ph-ph}$ the longer coupling to the environment (phonon-phonon) and $a, b$ are the relative weights as additional fit parameters .

The periodic oscillation of the AuNP size due to an excited breathing oscillation leads to a periodic red-shift of the plasmon absorption because of the variation in electron density (constant number of electrons for varying AuNP volume). In TA, the breathing mode is observable at the long-wavelength shoulder of the plasmon absorption, where the oscillation of the TA signal is superimposed with the contrast due to electron cooling. Here, we analyzed the breathing mode by probing $\Delta$A at 545 nm.

## Transient small-angle x-ray scattering

The experiment was conducted using the CAMP end-station at beamline BL1 of the FLASH facility at DESY,[5] extended by a user-provided injection system.[6,7] The free-electron laser (FEL) provided bursts of X-ray pulses with a photon energy of 275.5 eV and an intra-train repetition rate of 250 kHz. Each pulse train consisted of 100 pulses and the average fluence per pulse at the focus position was approximately $6 \cdot 10^9$ photons/mm$^2$. From indirect measurement of the electron pulse duration before the undulators, we estimate an FEL pulse duration of 120 fs. The optical pump pulses were provided by a chirped pulse optical parametric amplifier (OPCPA)[8] with an output wavelength of $(402.0 \pm 3.3)$ nm, a pulse duration of 80 fs and a repetition rate of 500 kHz. The pump pulses excited individual AuNPs and the scattering of time-delayed x-ray pulses was measured by large-area pnCCD photon detectors read-out at 10 Hz.[9,10]

Prior to the measurements, the spatial overlap of the injected particles, the FEL and the pump laser was established by focusing the FEL, the pump laser, and a frequency doubled Nd:YAG laser for particle detection on the same position of a retractable YAG screen.[11] The temporal overlap of the FEL and pump pulses was ensured by measuring the transient reflectivity of a Si$_3$N$_4$ film that was moved into the focus position, with x-ray excitation leading to an increased optical reflectivity of the Si$_3$N$_4$ film[12] for the pump pulses. Our data suggests a $< 500$ fs temporal resolution of the experiment. After retracting the screens, the particle beam was adjusted by maximizing scattering

of the particle-detection laser by injected sucrose particles and 80 nm AuNPs.

For sample injection, an AuNP aerosol was created using a commercial electrospray (TSI Advanced Electrospray 3482). Two differential pumping stages were used to remove excess nitrogen and $CO_2$ and the AuNP beam was generated using an optimized aerodynamic lens stack (ALS).[6,7,13] The injector pressure was kept at 1.0 mbar. The injector was tested prior to x-ray imaging by optical scattering off sucrose particles using a frequency doubled Nd:YAG laser (Innolas SpitLight) and a camera-based microscope system. The particle detection laser was turned off during the imaging experiments.

## Data analysis of transient small-angle x-ray scattering patterns

### Classification with *Dragonfly*

In an aerosol serial diffraction experiment, most XFEL pulses do not intercept a particle in the focus. Those that do, commonly termed 'hits', are detected by a straightforward total signal threshold on the diffracted data. But not all hits are useful since the aerosolization process produces signal from many objects other than spherical AuNPs. This includes multi-particle aggregates, ellipsoidal particles as well as patterns resulting from detector artifacts or cosmic rays. In order to extract the small signal from these patterns, one must find and reject these other hits. This classification was performed using the *Dragonfly*[14] software which uses a modified version of the EMC algorithm[15] in order to statistically align and average similar patterns into multiple class averages. The procedure was similar to that described in detail in[16] for classifying patterns from faceted AuNPs while taking into account in-plane rotation and detector panel gaps and bad pixels.

Figure S3 shows some representative class averages. We removed frames that were classified as contamination and multiple hits (top left and right panels) as well as ellipsoidal particles (bottom right panel) and only kept the single hits frames of spherical particles (bottom left panel).

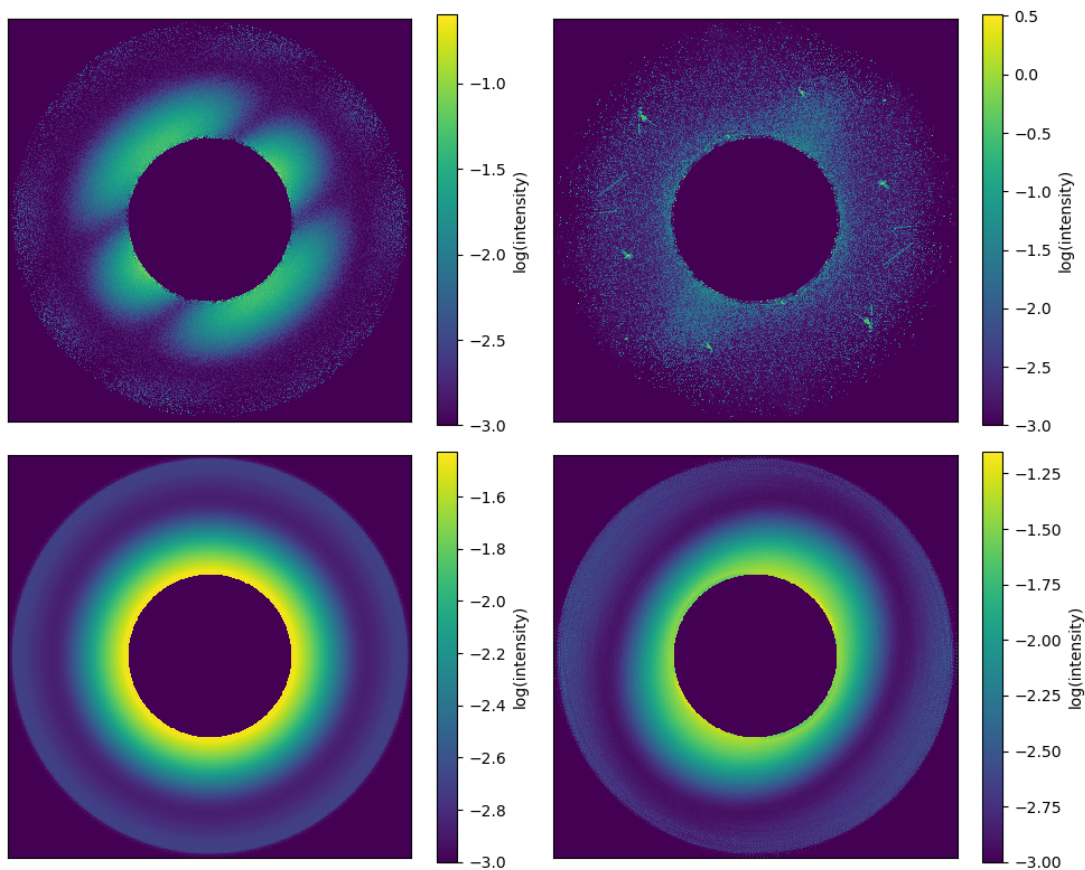

Fig. S3: **Four typical patterns from Dragonfly classifications.** Top left: multiple hits; top right: contaminations; bottom left: single hits of spherical AuNPs; bottom right: single hits of ellipsoidal particles.

**Diameter fitting**

All 2D class averages of single particles were fit with a function representing the diffraction from an ellipsoidal object. This resulted in three parameters per average, the major and minor axes as well as the tilt of the major axis. Only patterns belonging to class averages with a low eccentricity were included in further analysis. Since the EMC algorithm assigns each pattern a probability distribution over the various models, one can determine each AuNP's diameter by taking the weighted mean of the model diameters, with the weights obtained from the probabilistic assignment made by *Dragonfly*.

**Average diameter vs pump-probe delay**

In order to avoid experimental drifts, data was collected by continuously oscillating the pump-probe delay between -3 ps and +22 ps at a rate of 0.1 ps/s. Thus, in general, each measurement was obtained from a random delay time in this range. In order to aggregate the mean diameter at a given central delay time, the diameters of all patterns with a delay $\pm$ 1 ps from the central time were averaged. This bin width of 2 ps was chosen to balance the needs of observing the details of the oscillatory behavior with obtaining sufficient statistics. The bin centers were sampled every 0.1 ps in order to generate the data shown in Fig. 3 and Fig. 4a in the main text.

# Theory of electric and thermal sources to the oscillation

Our theoretical framework is based on a Heisenberg equation of motion framework that allows one to derive the AuNP size oscillation from the following Hamiltonian:

$$H = \sum_{\mathbf{k},\lambda} \epsilon_{\mathbf{k}}^{\lambda} \lambda_{\mathbf{k}}^{\dagger} \lambda_{\mathbf{k}} + \sum_{\mathbf{q}} \hbar\omega_{\mathbf{q}} b_{\mathbf{q}}^{\dagger} b_{\mathbf{q}} - \sum_{\mathbf{k},\mathbf{K},\lambda} \mathbf{d}^{\lambda\bar{\lambda}} \cdot \mathbf{E}_{-\mathbf{K}} \lambda_{\mathbf{k}}^{\dagger} \bar{\lambda}_{\mathbf{k}+\mathbf{K}} \qquad (1)$$

$$+ \sum_{\mathbf{q},\mathbf{k},\lambda} g_{\mathbf{q}}^{\lambda} \lambda_{\mathbf{k}+\mathbf{q}}^{\dagger} \lambda_{\mathbf{k}} \left( b_{\mathbf{q}} + b_{-\mathbf{q}}^{\dagger} \right) + \frac{1}{2} \sum_{\substack{\mathbf{k},\mathbf{k}',\mathbf{q} \\ \lambda\lambda'}} V_{\mathbf{q}} \lambda_{\mathbf{k}+\mathbf{q}}^{\dagger} \lambda_{\mathbf{k}'-\mathbf{q}}^{'\dagger} \lambda_{\mathbf{k}'}^{'} \lambda_{\mathbf{k}}$$

$$+ \sum_{\mathbf{q}_1 \mathbf{q}_2 \mathbf{q}_3} h_{\mathbf{q}_1 \mathbf{q}_2 \mathbf{q}_3} \left( b_{\mathbf{q}_1} + b_{-\mathbf{q}_1}^{\dagger} \right) \left( b_{\mathbf{q}_2} + b_{-\mathbf{q}_2}^{\dagger} \right) \left( b_{\mathbf{q}_3} + b_{-\mathbf{q}_3}^{\dagger} \right).$$

The first term accounts for the electron dispersion $\epsilon_{\mathbf{k}}^{\lambda}$ in the AuNP with electronic annihilation (creation) operators $\lambda_{\mathbf{k}}^{(\dagger)}$ with momentum $\mathbf{k}$ in the band $\lambda$, that allows one to treat the combined effect of intraband and interband transitions coupled to acoustic phonons.[17,18] The second term incorporates the dispersion $\hbar\omega_{\mathbf{q}}$ of dominant longitudinal acoustic (LA) phonon modes with phonon annihilation (creation) operators $b_{\mathbf{q}}^{(\dagger)}$ with momentum $\mathbf{q}$. In the Debye approximation,[19,20] the appearing phonon dispersion is assumed to be linear $\omega_{\mathbf{q}} = c_{LA}|\mathbf{q}|$ with the velocity of sound $c_{LA}$.[21] The third term describes the interband part of the semiclassical light-matter coupling[22,23] with the transition dipole matrix element $\mathbf{d}^{\lambda\bar{\lambda}} = 1/\Omega \int_{UC} u_{\lambda}^{*}(\mathbf{r}) \, e\mathbf{r} \, u_{\bar{\lambda}}(\mathbf{r}) \, d^3 r$ and the Fourier component $\mathbf{E}_{\mathbf{K}}(t)$ of the exciting electric field $\mathbf{E}(\mathbf{r},t) = \sum_{\mathbf{K}} \exp\{i\mathbf{K}\cdot\mathbf{r}\}\mathbf{E}_{\mathbf{K}}(t)$. As the experiment is carried out in a spectral region where interband transitions dominate the material response (cf.[24]), only the interband light-matter coupling will be considered in the following. The first term in the second line considers the electron-phonon interaction with the electron-phonon coupling strength $g_{\mathbf{q}}^{\lambda} = -i\sqrt{\hbar N/2M\omega_{\mathbf{q}}} \left[ \mathbf{A}_{\mathbf{q}} \cdot \mathbf{q} \right] \Phi_{\mathbf{q}}^{\lambda}$,[20] where $M$ is the ion mass in the unit cell, $N$ is the ion number in the crystal, and $\mathbf{A}_{\mathbf{q}}$ are the phonon polarization vectors given by $\mathbf{A}_{\mathbf{q}} = \frac{\mathbf{q}}{|\mathbf{q}|}$ for the case of $LA$ phonons.[25] $\Phi_{\mathbf{q}}^{\lambda}$ is the Fourier transformed ion electron-ion potential for the respective band $\lambda$.[26] The second term in that line accounts for the carrier-carrier Coulomb interaction $V_q = e^2/\epsilon(\mathbf{q})Vq^2$ with momentum exchange $\mathbf{q}$ between two carrier with momenta $\mathbf{k}$ and $\mathbf{k}'$ and the crystal volume

$V$. The third line describes the phonon-phonon interaction occurring from anharmonic corrections in the ion-ion interaction. The appearing matrix element conserves the total momentum of the involved phonons, i.e. $h_{\mathbf{q_1 q_2 q_3}} = \tilde{h}_{\mathbf{q_1 q_2 q_3}} \delta_{\mathbf{q_1+q_2+q_3},0}$.

We define the microscopic phonon mode amplitude $s_\mathbf{q}$ that determines the macroscopic lattice displacement (Eq. (1) in the main document) and the Wigner functions for the band occupation $f_\mathbf{k}^\lambda(\mathbf{r})$ and the interband coherence $p_\mathbf{k}(\mathbf{r})$ for an effective two-band model with a derived susceptibility that can be obtained from the experiment:

$$s_\mathbf{q}(t) = \frac{1}{2}\left(\langle b_\mathbf{q}\rangle(t) + \langle b^\dagger_{-\mathbf{q}}\rangle(t)\right), \tag{2}$$

$$f_\mathbf{k}^\lambda(\mathbf{r},t) = \sum_\mathbf{q} e^{i\mathbf{q}\cdot\mathbf{r}} \langle \lambda^\dagger_{\mathbf{k-q}} \lambda_\mathbf{k}\rangle(t), \tag{3}$$

$$p_\mathbf{k}(\mathbf{r},t) = \sum_\mathbf{q} e^{i\mathbf{q}\cdot\mathbf{r}} \langle v^\dagger_{\mathbf{k-q}} c_\mathbf{k}\rangle(t). \tag{4}$$

In gold, the interband transitions occur between the initially occupied and the initially unoccupied band[27] that are below (above) the Fermi level respectively. Accordingly, they are labeled as v for valence and c for conduction band. The equation of motion for the phonon mode amplitude reads:

$$\left(\partial_t^2 + 2\gamma_{\text{ph}}^\mathbf{q} \partial_t + \omega_\mathbf{q}^2\right) s_\mathbf{q}(t) = -\frac{\omega_\mathbf{q}}{\hbar} \sum_\mathbf{k} \left(g^c_{-\mathbf{q}} - g^v_{-\mathbf{q}}\right) \tilde{f}_\mathbf{k}^c(\mathbf{q},t) \tag{5}$$

$$-\frac{3\omega_\mathbf{q}}{\hbar} \sum_\mathbf{k} \tilde{h}_{-\mathbf{q},\mathbf{k},\mathbf{q-k}}\, \delta\tilde{n}_\mathbf{k}(\mathbf{q},t). \tag{6}$$

This equation describes the dynamics of a classical damped oscillator equation with mode index $\mathbf{q}$. The left-hand side describes the oscillation of the coherent phonon amplitude $s_\mathbf{q}$ with a damping rate $\gamma_{\text{ph}}^\mathbf{q}$ [28] resulting from phonon-phonon interactions, treated here as a constant. On the right-hand side, two sources to the oscillation of the coherent phonon mode amplitude are identified. The first term is determined by the dynamics of the Fourier transform of the occupation Wigner function $\tilde{f}_\mathbf{k}^\lambda(\mathbf{q}) = \langle \lambda^\dagger_{\mathbf{k-q}} \lambda_\mathbf{k}\rangle$ showing that a spatial inhomogeneous electron distribution drives $s_\mathbf{q}$.[29]

11

The prefactor scales with the difference $(g^c_{-\mathbf{q}} - g^v_{-\mathbf{q}})$ of the electron-phonon coupling element of conduction and valence band. The second source of the coherent phonon amplitude results from thermal effects of the change of the phonon occupation caused by electron heating from its equilibrium value: $\delta \tilde{n}_\mathbf{k}(\mathbf{q},t) = \tilde{n}_\mathbf{k}(\mathbf{q},t) - \tilde{n}_\mathbf{k}(\mathbf{q}, t \to -\infty)$ where we define $\tilde{n}_\mathbf{k}(\mathbf{q},t) = \langle b^\dagger_{\mathbf{k}-\mathbf{q}} b_\mathbf{k} \rangle (t)$. The constant contributions from the right side (equilibrium electron density and equilibrium phonon density) were absorbed in the equilibrium position of the oscillator, so that only deviations from equilibrium act as sources for the oscillation.

As the Wigner occupations $f^\lambda_\mathbf{k}(\mathbf{r})$ act as sources of the coherent phonon amplitude Eq. (5), we also derive an equation of motion for a two band model within the gradient expansion[29] and obtain

$$\partial_t f^v_\mathbf{k}(\mathbf{r}) = -\mathbf{v}^v_\mathbf{k} \cdot \nabla_\mathbf{r} f^v_\mathbf{k}(\mathbf{r}) - 2\,\mathrm{Im}\left\{ \frac{\mathbf{d}^{vc} \cdot \mathbf{E}(\mathbf{r},t)}{\hbar} p_\mathbf{k}(\mathbf{r}) \right\} \quad (7)$$

$$\partial_t f^c_\mathbf{k}(\mathbf{r}) = -\mathbf{v}^c_\mathbf{k} \cdot \nabla_\mathbf{r} f^c_\mathbf{k}(\mathbf{r}) + 2\,\mathrm{Im}\left\{ \frac{\mathbf{d}^{vc} \cdot \mathbf{E}(\mathbf{r},t)}{\hbar} p_\mathbf{k}(\mathbf{r}) \right\} \quad (8)$$

$$\partial_t p_\mathbf{k}(\mathbf{r}) = \left[ -\frac{i}{\hbar}(\epsilon^c_\mathbf{k} - \epsilon^v_\mathbf{k}) - \gamma - \frac{(\mathbf{v}^c_\mathbf{k} + \mathbf{v}^v_\mathbf{k})}{2} \cdot \nabla_\mathbf{r} \right] p_\mathbf{k}(\mathbf{r})$$
$$+ i\frac{\mathbf{d}^{cv} \cdot \mathbf{E}(\mathbf{r},t)}{h}[f^v_\mathbf{k}(\mathbf{r}) - f^c_\mathbf{k}(\mathbf{r})] \quad (9)$$

In Eqs. (7,8), we identify the group velocity $\mathbf{v}^\lambda_\mathbf{k} = \nabla_\mathbf{k} \epsilon^\lambda_\mathbf{k}/\hbar$ for the band $\lambda$ that considers the drift of the electronic Wigner occupation. The term on the right-hand side in Eqs. (7,8) accounts for the optical source via the full electric field $\mathbf{E}(\mathbf{r},t)$ that includes the external field as well as polarization contributions. In Eq. (9), the interband polarization oscillates with the band gap $(\epsilon^c_\mathbf{k} - \epsilon^v_\mathbf{k})$ and is also driven by the external optical driving field $\mathbf{E}(\mathbf{r},t)$. $\gamma$ is the electron-electron scattering induced damping term of the order of few femtoseconds[30] for the interband transition and was added phenomenologically.

In the following we expand the individual Wigner functions in orders of the electric field $f^\lambda_\mathbf{k}(\mathbf{r},t) = f^{0,\lambda}_\mathbf{k}(\mathbf{r}) + f^{1,\lambda}_\mathbf{k}(\mathbf{r},t) + f^{2,\lambda}_\mathbf{k}(\mathbf{r},t) + \mathcal{O}(\mathbf{E}^3)$ and $p_\mathbf{k}(\mathbf{r},t) = p^1_\mathbf{k}(\mathbf{r},t) + p^2_\mathbf{k}(\mathbf{r},t) + \mathcal{O}(\mathbf{E}^3)$, enter a rotating frame $\mathbf{E}(\mathbf{r},t) = \tilde{\mathbf{E}}^+ e^{i\omega_\mathrm{opt} t} + \tilde{\mathbf{E}}^- e^{-i\omega_\mathrm{opt} t}$ and $p_\mathbf{k}(\mathbf{r},t) = \tilde{p}_\mathbf{k}(\mathbf{r},t)e^{-i\omega_\mathrm{opt} t}$ and apply a rotating wave

12

approximation. Assuming a fast dephasing $\gamma$, we solve the equation for the interband coherence adiabatically and also neglect the transport terms in Eqs. (9,8,7) as they are also small compared the individual time dynamics. Since the spatial gradients needed to evaluate the displacement Eq. (16) is mainly determined by the dielectric, off-resonant contribution, we replace the full self-consistent field by the field $\tilde{\mathbf{E}}_B$ formed by the dielectric background. In addition, a radiation self energy correction will be absorbed in the frequency $\omega_\mathbf{k} \equiv (\epsilon^c_\mathbf{k} - \epsilon^v_\mathbf{k})\hbar \to \tilde{\omega}_\mathbf{k}$ and the dephasing $\gamma \to \tilde{\gamma}$. Thus, the governing equations for the band occupations read

$$\partial_t f^{c,2}_\mathbf{k}(\mathbf{r},t) = \left|\frac{\mathbf{d}^{vc} \cdot \tilde{\mathbf{E}}_B(\mathbf{r},t)}{\hbar}\right|^2 \frac{2\tilde{\gamma}}{\tilde{\gamma}^2 + (\tilde{\omega}_\mathbf{k} - \omega_\text{opt})^2} \left[f^{v,0}_\mathbf{k}(\mathbf{r}) - f^{c,0}_\mathbf{k}(\mathbf{r})\right], \tag{10}$$

$$\partial_t f^{v,2}_\mathbf{k}(\mathbf{r},t) = -\left|\frac{\mathbf{d}^{vc} \cdot \tilde{\mathbf{E}}_B(\mathbf{r},t)}{\hbar}\right|^2 \frac{2\tilde{\gamma}}{\tilde{\gamma}^2 + (\tilde{\omega}_\mathbf{k} - \omega_\text{opt})^2} \left[f^{v,0}_\mathbf{k}(\mathbf{r}) - f^{c,0}_\mathbf{k}(\mathbf{r})\right]. \tag{11}$$

In the following, we apply a coarse-graining procedure using the macroscopic definition of the lattice displacement (Eq. 1 in the main manuscript) and a momentum expansion for the Wigner occupations given by:

$$\rho^\lambda(\mathbf{r},t) \equiv \frac{e}{\Omega} \sum_\mathbf{k} f^\lambda_\mathbf{k}(\mathbf{r},t), \tag{12}$$

with the unit cell volume $\Omega$. This allows to identify the contribution of the interband transitions via the macroscopic susceptibility:

$$\chi^\text{inter}(\omega) = -\frac{|d|^2}{\hbar\epsilon_0\Omega} \sum_\mathbf{k} \frac{f^{v,0}_\mathbf{k}(\mathbf{r}) - f^{c,0}_\mathbf{k}(\mathbf{r})}{\omega - \tilde{\omega}_\mathbf{k} + i\tilde{\gamma}}, \tag{13}$$

resulting in macroscopic equations for the carrier densities:

$$\partial_t \rho^v_2(\mathbf{r},t) = -\frac{2e\epsilon_0}{\hbar}\left|\tilde{\mathbf{E}}_B(\mathbf{r},t)\right|^2 \text{Im}\{\chi^\text{inter}(\omega_\text{opt})\}, \tag{14}$$

$$\partial_t \rho^c_2(\mathbf{r},t) = \frac{2e\epsilon_0}{\hbar}\left|\tilde{\mathbf{E}}_B(\mathbf{r},t)\right|^2 \text{Im}\{\chi^\text{inter}(\omega_\text{opt})\}. \tag{15}$$

These equations describe the occuption dynamics for the individual bands with a source term that creates or annihilates density in the respective band due to the presence of an external field $\tilde{\mathbf{E}}_0(\mathbf{r}, t)$.

Using the definition of the lattice displacement $\mathbf{u}(\mathbf{r}, t)$, the equation for the lattice displacement can be found:

$$\left[\partial_t^2 + 2\gamma_{\text{ph}}\partial_t - c_{LA}^2 \nabla_\mathbf{r}^2\right] \mathbf{u}(\mathbf{r}, t) = \zeta \, \nabla \rho_2^c(\mathbf{r}, t) + \boldsymbol{\xi} \left[T(t) - T_0\right]. \tag{16}$$

The temperature dependence is obtained within a Debye approximation for the incoherent phonon distribution. We find two driving terms with prefactors $\zeta$ and $\boldsymbol{\xi}$ that characterize the macroscopic model. The first represents the newly found displacement source resulting from an optically induced gradient in the spatial electron distribution $\rho_2^c(\mathbf{r}, t)$. This spatial gradient in the electron density is identified as the dominating origin of the optically induced oscillation of the displacement since it can explain the onset of the oscillation, measured to be faster than the thermal contributions (second term) in Eq. (16). The prefactor $\zeta$ scales with the difference of the electron-phonon coupling elements of the conduction and valence bands. The difference of the two is fitted to the experimental data as 8 % of the individual valence band electron-phonon coupling elements. This is on the same order of magnitude as the estimations in[31] which are based on DOS averaged *ab initio* calculations.

The second term occurs via the temperature difference as source and is found in agreement with previous thermal models (cf.[28,32]). The thermal source acts via the lattice temperature change and causes an additional thermal expansion of the lattice which also impulsively excites oscillations of the nanoparticle and a shift in the equilibrium position particle radius. Within our model, the temperature changes were modeled using a two temperature model.[28,32]

In order to solve the set of coupled partial differential equations they will be mapped on so-called Lamb modes $\mathbf{u}_{\text{nlm}}(\mathbf{r}) = \nabla \phi_{\text{nlm}}(\mathbf{r})$ with $\phi_{\text{nlm}}(\mathbf{r}) = j_l(\alpha_n r) P_l^m(\cos\theta) \exp\{im\varphi\}$ that character-

14

ize the vibrational modes of the a sphere.[33] As under the stress-free boundary assumptions those modes are not orthogonal, we expand the dynamics in the vibrational ground mode of the nanoparticle only. Hence, the expansion reads $\mathbf{u}(\mathbf{r},t) = a(t)\,\mathbf{u}_0(\mathbf{r}), \rho_2^c(\mathbf{r},t) = \rho(t)\,\phi_0(\mathbf{r})$ and the overlap normalization factor is $A = \int_{V_s} \phi_0^*(\mathbf{r})\phi_0(\mathbf{r})\mathrm{d}^3r$ with the volume of the nanoparticle $V_s$. Using the projection, the equations simplify to ordinary differential equations which can numerically be integrated:

$$\partial_t \rho_2^c(t) = \frac{2e\epsilon_0}{\hbar A} \operatorname{Im}\{\chi^{\text{inter}}(\omega_{\text{opt}})\} \int_{V_s} \mathrm{d}^3 r \, \mathbf{u}_0^*(\mathbf{r}) \cdot \nabla \left|\tilde{\mathbf{E}}_B(\mathbf{r},t)\right|^2, \tag{17}$$

$$\left[\partial_t^2 + 2\gamma_{\text{ph}}\partial_t + \omega_0^2\right] a(t) = \zeta \rho_2^c(t) + \xi\left[T(t) - T_0\right]. \tag{18}$$

$\epsilon_d$ is the intraband Drude response of the background that is not incorporated in the interband susceptibility $\chi^{\text{inter}}$. In order to compare with the experimental observations the relative oscillation $\mathbf{u}(\mathbf{R},t)/|\mathbf{R}|$ where $\mathbf{R}$ is chosen to be on the boundary of the sphere. Our numerical implementation allows to artificially switch the driving sources on and off which will allow to study their respective influence separately for various parameter regimes.

## Two-temperature model

The energy conversion within AuNPs by electron-phonon coupling is typically modeled in terms of the electron and lattice temperatures ($T_e/T_l$). Two coupled differential equations describe the change of the two temperatures with time:

$$C_e(T_e)\frac{\partial T_e}{\partial t} = -g(T_e - T_l) + \frac{W_0}{\sqrt{\pi}\sigma}\exp\{-t^2/\sigma^2\}, \qquad (19)$$

$$C_l\frac{\partial T_l}{\partial t} = g(T_e - T_l) - (T_l - 298)/\tau_s, \qquad (20)$$

where $C_e$ and $C_L$ are the electron and lattice heat capacities, respectively, $g$ is the electron-phonon coupling constant. The last term in Eq. (19) is the source of hot electron excitation with the amplitude $W_0$ and width $\sigma$ of the optical pulse. $\tau_s$ in Eq. (20) is the time constant for heat transfer from the phonon subsystem to the surrounding medium.[34,35]

To simulate the electron and lattice temperature numerically using the two-temperature model, the lattice temperature was set to room temperature, while the initial electron temperature had to be calculated as follows: The initial jump in electronic temperature $\Delta T_{\text{el}}^{\text{initial}}$ depends on the absorbed energy per pump pulse and unit volume of Au:

$$\frac{I_{\text{abs}}}{N_{\text{AuNP}} \cdot V_{\text{AuNP}}} = \frac{1}{2}\gamma((\Delta T_{\text{el}}^{\text{initial}} + T_{\text{RT}})^2 + (T_{\text{RT}})^2), \qquad (21)$$

where $I_{\text{abs}}$ is the absorbed energy per pump pulse according to Lambert-Beer's Law, $N_{\text{AuNP}}$ is the number of AuNPs in the pump beam, $V_{\text{AuNP}}$ is the volume of a AuNP of a given size, $\gamma$ is the electron heat capacity for bulk gold and $T_{\text{RT}}$ is the room temperature.[36]

In this work, the absorbed energy per unit volume of Au was calculated based on the concentration of the particle solution and the absorbance determined by UV-Vis spectroscopy and the excitation power measured during the TA experiments. The e-ph coupling constant was taken from our previous work[37] and the electron and lattice heat capacities were taken from literature.[31,38]



\end{document}